\documentclass[a4paper,11pt]{article}
\usepackage{pos}
\usepackage{slashed}
\usepackage{dsfont}

\usepackage{tikz}
\usepackage{graphicx}

\title{Resummation of small-spin singularities in anomalous dimensions of twist-two operators}
\ShortTitle{Small-spin singularities in anomalous dimensions of twist-two operators}

\author[a,b]{Alexander N. Manashov}
\author[a]{Sven-Olaf Moch}
\author*[a]{Leonid A. Shumilov}

\affiliation[a]{II. Institut für Theoretische Physik, Universität Hamburg,\\
  Luruper Chaussee 149, D-22761 Hamburg, Germany}

\affiliation[b]{Institut für Theoretische Physik, Universität Regensburg,\\
Universitätsstrasse 31, D-93040 Regensburg, Germany}

\emailAdd{alexander.manashov@desy.de}
\emailAdd{sven-olaf.moch@desy.de}
\emailAdd{leonid.shumilov@desy.de}

\abstract{Anomalous dimensions of leading-twist operators in QCD play an important role in precision predictions for high-energy processes, since they govern the scale evolution of parton distributions. Their analytic structure as a function of spin is particularly important due to the complexity of higher-loop computations. In these proceedings, we discuss the resummation of the certain type of such singularities that share common features with those appearing in the quark flavor-nonsinglet sector of QCD. Our main focus is on the interplay between Gross-Neveu-Yukawa model in $\epsilon$ expansion and Gross-Neveu in $1/N$ expansion. Such resummation allows one to predict the higher-loop singular behavior and reveals connections with the conformal Regge theory and recent studies of detector operators in QCD and various conformal field theories.}

\FullConference{17th International Symposium on Radiative Corrections: Applications of Quantum Field Theory to Phenomenology (RADCOR2025)\\
5-10 October 2025\\
Puri, India\\}

\begin{document}
\maketitle

\section{Introduction}

The main objects of interest in these proceedings are leading-twist operators. In the case of the flavor non-singlet quark operator in QCD, it has the following form
\begin{equation}
\label{QCD-op-def}
\mathcal{O}^{q}_{\mu_1\ldots\mu_s}(x) = \bar{q}(x)\gamma_{(\mu_1}D_{\mu_2}\ldots D_{\mu_s)}q(x) - \text{traces},
\end{equation}
where $D_{\mu}$ is the covariant derivative, and the quark fields $\bar{q}(x)$ and $q(x)$ are assumed to have different flavors. The tensor structure is chosen in such a way that the operator has spin $s$. The quantity naturally associated with the operator~\eqref{QCD-op-def} is its anomalous dimension. It measures the violation from canonical scaling in the interacting theory and admits a perturbative expansion
\begin{equation}
\label{gamma-expansion}
\gamma_{\text{ns}}(s) = a\gamma_{\text{ns}}^{(1)}(s) + a^2\gamma_{\text{ns}}^{(2)}(s) + \ldots,
\end{equation}
where $a = \alpha_s/4\pi$. In QCD, such anomalous dimensions control the scale dependence of parton distribution functions via the Wilson OPE, and therefore have direct phenomenological importance. The analogous quantities are also interesting in CFTs, as they determine the leading contributions to the short-distance expansion of correlation functions.

Computing the perturbative coefficients in~\eqref{gamma-expansion} is quite involved, especially at higher-loop orders. The state-of-the-art prediction is now at three loops~\cite{Moch:2004pa} (see also~\cite{Vogt:2004mw} for the singlet case), while only partial results are available at four loops (see, for example~\cite{Moch:2017uml, Gehrmann:2023iah, Falcioni:2024qpd}). The natural way of organizing such calculations typically provides results for specific values of $s \in \mathbb{Z}_{>0}$, and the complexity of calculations grows very fast with $s$. A recent update pushed this to $s = 22$~\cite{Falcioni:2025hfz}. The desired result, however, is to obtain $\gamma_{\text{ns}}^{(\ell)}(s)$ as a function of $s$ based on this available data. Obviously, a few points are not enough to fully constrain the functional form, but surprisingly this problem can be effectively solved by considering a set of Diophantine equations (see, e.g., the reconstruction of the three-loop transversity result in~\cite{Velizhanin:2012nm}). In the case of the operator~\eqref{QCD-op-def}, such a reconstruction was performed in the planar limit in~\cite{Moch:2017uml}, as well as for the $\zeta$-~\cite{Kniehl:2025ttz} and fermionic contributions~\cite{Kniehl:2025jfs}.

The efficiency of such an approach increases once more constraints on the functional form of $\gamma_{\text{ns}}^{(\ell)}(s)$ are available. For example, valuable information comes from fixing the large-$s$ behavior~\cite{Basso:2006nk, Dokshitzer:2006nm}, which allows one to express everything in terms of the so-called reciprocity-respecting functions~\cite{Beccaria:2010tb}. In these proceedings, we focus instead on the limit of small spins, namely $s \to 0$, for the operators~\eqref{QCD-op-def}. The anomalous dimensions in this limit become singular, and we discuss how such singularities can be resummed according to the procedure introduced in~\cite{Caron-Huot:2022eqs}. We review this technique in the $O(N)$-symmetric $\varphi^4$ model and then apply the same resummation strategy to the Gross–Neveu–Yukawa model in the $\epsilon$ expansion and to the Gross–Neveu model in the $1/N$ expansion, studying the conformal Regge trajectories in these theories. We focus on these models because the limit $s \to 0$ shares common features with the corresponding behavior of the anomalous dimensions of the operator~\eqref{QCD-op-def}. Note also the work~\cite{Li:2025knf}, where a related one-loop analysis was carried out in QCD, as well as in the case of the $s \to 1$ singularities in the gluon-gluon anomalous dimension.

\section{The structure of singularities}

Let us now discuss in more detail the singularities that appear in the anomalous dimensions of the operators~\eqref{QCD-op-def}. In the limit $s \to 0$ one expects a specific behavior according to the well-studied functional form of known results. In particular, we expect the anomalous dimension at $\ell$-th loop order, $\gamma^{(\ell)}(s)$, to consist of harmonic sums~\cite{Vermaseren:1998uu} of the transcendental weight $\le 2\ell - 1$, multiplied by rational functions of $s$. While we do not focus on the precise structure of the latter, general arguments imply that in the limit $s \to 0$, anomalous dimensions behave as
\begin{equation}
\label{sing-exp}
    \gamma^{(\ell)}_{\text{ns}}(s) = \dfrac{\#}{s^{2\ell - 1}} + \dfrac{\#}{s^{2\ell - 2}} + O\left(\dfrac{1}{s^{2\ell - 3}}\right).
\end{equation}
To be rigorous we note that studying the continuous limit in $\gamma_{\text{ns}}^{(\ell)}(s)$ first requires analytical continuation from $s \in \mathbb{Z}_{>0}$. Using known functional form, the domain can be naturally extended to $s \in \mathbb{C}\backslash\{0, -1, -2, \ldots\}$: for the rational functions this continuation is straightforward, and for the harmonic sums it is also well-known.

The presence of a singularity at $s \to 0$ at first sight does not seem contradictory, $s = 0$ in the definition~\eqref{QCD-op-def} does not correspond to any local operator. Nevertheless, going beyond fixed order perturbation theory, it is natural to assume that singular contribution may resum to a function regular at this point. The first attempt in this direction was made in~\cite{Ermolaev:1995fx, Blumlein:1995jp}, where $\gamma_{\text{ns}}^{(\ell)}(s)$ was analyzed as a solution of a Bethe-Salpeter equation. As a result, the so-called double-logarithmic equation (DLE) was derived
\begin{equation}
\label{dle-1loop}
    2\gamma_{\text{ns}}(s)\left(s + \dfrac{1}{2}\gamma_{\text{ns}}(s)\right) = -8C_Fa.
\end{equation}
Solving this quadratic equation, one acquires the all-order prediction for the leading singularities at each order
\begin{equation}
\label{non-singlet-resum}
    \gamma_{\text{ns}}(s) = -s + \sqrt{s^2 - 8C_Fa} = -4\dfrac{C_Fa}{s} - 8\dfrac{C_F^2a^2}{s^2} - 32\dfrac{C_F^3a^3}{s^3} + O\left(a^4\right).
\end{equation}

Although the DLE derivation is explicitly based on one-loop result, the equation~\eqref{dle-1loop} turns out to have a much wider range of validity. It was noticed in~\cite{Velizhanin:2011pb} that replacing $\gamma_{\text{ns}}(s)$ in~\eqref{dle-1loop} by the corresponding anomalous dimension in planar $\mathcal{N} = 4$ super Yang-Mills produce the regular function on the r.h.s. in all known loop orders. The situation starts to break in the non-planar sector starting at four loops, however, eq.~\eqref{dle-1loop} still partially resums subleading singularities. Moreover, in~\cite{Velizhanin:2014dia} it was observed that a simple modification of~\eqref{dle-1loop}
\begin{equation}
\label{dle-qcd}
    \gamma_{\text{ns}}(s)\left(s + \bar{\beta}(a) + \dfrac{1}{2}\gamma_{\text{ns}}(s)\right),
\end{equation}
where $\beta(a) = -2a(\epsilon + \bar{\beta}(a))$, allows one to resum all singular contributions in the planar limit at every known loop order. For the non-planar terms, subleading singularities start to appear at three loops. Despite the remarkable empirical success of~\eqref{dle-qcd}, a theoretical explanation was missing. A suitable framework for the possible explanation was derived in the context of CFT~\cite{Caron-Huot:2022eqs} and relies on the studies of the analytical continuation of the CFT spectrum~\cite{Caron-Huot:2017vep} and leading-twist operators~\cite{Kravchuk:2018htv}. In the following section we briefly review this approach in the simple case of the critical $O(N)$-symmetric $\varphi^4$ theory. 

Before closing this section, let us mention a practical importance of~\eqref{dle-qcd}. The regularity of this expression can be seen as the set of constraints which allow to predict the singular structure of $\gamma_{\text{ns}}^{(\ell)}(s)$ based on the lower loop data. These predictions, in turn, can be effectively used in the program of deriving an analytical expression for $\gamma_{\text{ns}}^{(\ell)}(s)$ based on the finite number of points. 

\section{$O(N)$-symmetric $\varphi^4$ model \label{sect:phi4}}

This section is dedicated to the $O(N)$-symmetric $\varphi^4$ model
\begin{equation}
\label{phi4-action}
    S = \int d^dx\Big((\partial\varphi)^2 + \dfrac{g}{4!}\left(\varphi^2\right)^2\Big),
\end{equation}
where $\varphi^a(x)$ ($a = 1, \ldots, N$) transforms under the fundamental representation of the $O(N)$ group. We focus on the singlet twist-2 operator, which, in analogy with~\eqref{QCD-op-def}, is defined as~\footnote{See~\cite{Manashov:2025kgf} for the discussion of other types of twist-2 operators in this model.}
\begin{equation}
\label{phi4-op-def}
    \mathcal{O}_s(x) = \sum_{a = 1}^{N}\varphi^a(x)\partial_{\mu_1}\ldots\partial_{\mu_s}\varphi^a(x) - \text{traces}. 
\end{equation}
The anomalous dimension of this operator is known to four-loop accuracy~\cite{Derkachov:1997pf,Manashov:2017xtt} ($u = g^2/(4\pi)^2$)
\begin{equation}
\label{phi4-anomalous-dimension}
     \gamma(s) = 2\gamma_{\varphi} - u^2\dfrac{N + 2}{9s(s + 1)}\Bigg(3 + \dfrac{u}{3}(N + 8)\Bigg(S_1(s) - 5 + \dfrac{3}{2}\dfrac{2s + 1}{s(s + 1)}\Bigg)\Bigg) + O(u^4).
\end{equation}

The expression~\eqref{phi4-anomalous-dimension} possesses singularities at $s \in \mathbb{Z}_{\le 0}$. Compared to~\eqref{QCD-op-def} in QCD, the limit $s \to 0$ is even more intriguing here, because the local operator at this point is well-defined. From~\eqref{phi4-op-def} we obtain $\mathcal{O}_{s=0}=\varphi^2$, whose anomalous dimension reads 
\begin{equation}
    \gamma_{\varphi^2} = u\dfrac{N + 2}{3}\left(1 - \dfrac{5u}{6} + u^2\dfrac{5N + 37}{12}\right) + O(u^4).
\end{equation} 
Hence we encounter the obvious mismatch
\begin{equation}
\label{mismatch}
    \lim_{s \to 0}\gamma(s) = \infty \neq \gamma_{\varphi^2}.
\end{equation}
Remarkably, the same combination of anomalous dimensions~\eqref{dle-qcd} that appeared in QCD, turns out to be regular in all known orders of perturbation theory in this model. It therefore can be used to perform a resummation, solving the quadratic equation as in~\eqref{non-singlet-resum}, and this resummation procedure also solves the problem stated in~\eqref{mismatch}. We note that this fact is far from being obvious, since in $\varphi^4$ there is no direct analogue of the Bethe-Salpeter-equation-based derivation~\cite{Ermolaev:1995fx, Blumlein:1995jp}. 

The regularity of the combination~\eqref{dle-qcd} can be justified at the critical point, where the model~\eqref{phi4-action} describes a proper CFT. Assuming $d = 4 - 2\epsilon$, one can tune the coupling $u \mapsto u^*$, such as $\beta(u^*) = 0$. The expression for $u^* = u^*(\epsilon)$ can be found perturbatively, which makes $\epsilon$ a parameter of the perturbative expansion. Once we consider the theory at the critical point we can translate the consideration of the Wilson-Fisher CFT (the $N = 1$ case)~\cite{Caron-Huot:2022eqs} to arbitrary $N$ without changes. Note also the work~\cite{Chang:2025zib} where the consideration is done for the model~\eqref{phi4-action} in the $1/N$ expansion. 

The crucial point of staying in the CFT setup is that we can consider an analytical continuation in spin not only on the level of function~\eqref{phi4-anomalous-dimension}, but on the level of the CFT spectrum~\cite{Caron-Huot:2017vep}. In particular, at the point $s = 0$ the trajectory defined by the continuation of critical anomalous dimension, $\gamma^*(s)$~\footnote{Here and in what follows by star we denote the corresponding value at the critical coupling.}, mixes with the so-called shadow trajectory, $\widetilde{\gamma}^*(s)$. In more detail, we consider two operators: first, analytical continuation of~\eqref{phi4-op-def}, operator $\mathcal{O}_{\Delta(s)}$ with the scaling dimension $\Delta(s) = d - 2 + s + \gamma^*(s)$ and its shadow transform, $\mathcal{O}_{\widetilde{\Delta}(s)}$\footnote{For the integer values of $s$ this is a non-local operator, defined by the shadow transform, for details see~\cite{Kravchuk:2018htv}.}, with the scaling dimension $\widetilde{\Delta}(s) = d - \Delta(s)$. Both belong to the analytically continued spectrum and at the point $s = 0$ their scaling dimensions coincide in the free theory ($\epsilon = 0$)
\begin{equation}
\label{degeneracy}
    \Delta_0(0) = 2 = \widetilde{\Delta}_0(0).
\end{equation}

When the interaction is turned on ($\epsilon \neq 0$), guided~\footnote{Here we consider it only as a motivation. For more rigorous considerations one has to simultaneously renormalize the analytically continued operators $\mathcal{O}_{\Delta(s)}$ and $O_{\widetilde{\Delta}}(s)$ in a vicinity of $s = 0$.} by the von Neumann-Wigner non-crossing rule (see~\cite{Korchemsky:2015cyx} for consideration in a unitary CFT), we expect this degeneracy to be resolved. Moreover, we expect the exact trajectory to be a multi-valued function, regular at $s = 0$, which interpolates between $\Delta(s)$ and $\widetilde{\Delta}(s)$ upon moving between different sheets of the Riemann surface. Such a function can be naturally defined as a solution of an equation
\begin{equation}
\label{equation-reg}
    x^2 - (\widetilde{\Delta}(s) + \Delta(s))x + \widetilde{\Delta}(s)\cdot \Delta(s) = 0.
\end{equation}
The regularity at $s = 0$, in turn, imposes constraints on the coefficients in equation~\eqref{equation-reg}, namely
\begin{equation}
    \begin{cases}
        \widetilde{\Delta}(s) + \Delta(s) \text{ is regular at } s = 0, \\
        \widetilde{\Delta}(s)\cdot\Delta(s) ~\text{ is regular at } s = 0.
    \end{cases}
\end{equation}
The first constraint is trivial ($\widetilde{\Delta} + \Delta = d$), but the second one is not. Substituting $\Delta(s) = d - 2 + s + \gamma^*(s)$ in the product $\widetilde{\Delta}(s)\cdot\Delta(s)$ and omitting a priori regular terms, we get
\begin{equation}
\label{delta-m-def}
    \delta m_*^2(s) = 2\gamma^*(s)\left(s - \epsilon + \dfrac{1}{2}\gamma^*(s)\right),
\end{equation}
which must be regular at $s = 0$. This prediction is confirmed by perturbative calculations: it was verified in~\cite{Kravchuk:2018htv} for the case $N = 1$ and in~\cite{Manashov:2025kgf} for arbitrary $N$ and different types of twist-2 operators. 

Regularity of~\eqref{delta-m-def} leads to a resummation prescription. In the vicinity of $s = 0$, instead of the perturbative expansion~\eqref{phi4-anomalous-dimension}, we define the anomalous dimension as 
\begin{equation}
\label{resum-crit}
    \gamma^*(s) = - s + \epsilon + \sqrt{(s - \epsilon)^2 + \delta m_*^2(s)},
\end{equation}
where the square root is understood as a function defined on the appropriate Riemann surface. 

A remarkable property of the resummation formula~\eqref{resum-crit} is that it can be directly translated to arbitrary coupling. Following the logic presented in~\cite{Basso:2006nk} one substitutes $\epsilon = -\bar{\beta}(u^*)$, where $\beta(u) = -2u(\epsilon + \bar{\beta}(u))$, and then changes $u^* \mapsto u$ together with $\gamma^*(s) \mapsto \gamma(s)$. In $\overline{\text{MS}}$-like schemes this step is justified by the independence of perturbative coefficients of~\eqref{phi4-anomalous-dimension} from the space-time dimension. This gives the following rule for the resummation, which repeats the solution of the DLE~\eqref{dle-qcd}
\begin{equation}
\label{resum-arbitrary}
    \gamma(s) = -s -\bar{\beta}(u) + \sqrt{(s + \bar{\beta}(u))^2 + \delta m^2(s)}.
\end{equation}
One can explicitly check that on the appropriate branch the square root reproduces the expected behavior, namely, if $s$ is far from $s = 0$, \eqref{resum-arbitrary} gives~\eqref{phi4-anomalous-dimension}, and $\lim_{s \to 0}\gamma(s) = \gamma_{\varphi^2}$.

At the end of this section we would like to mention another remarkable property of combination~\eqref{delta-m-def}. In the critical three-dimensional $O(N)$-vector model the same combination of anomalous dimensions corresponds to the mass-corrections of the higher spin fields in the dual $\text{AdS}_4$ description~\cite{Klebanov:2002ja}. It would be highly
desirable to gain a better understanding of the role of $\text{AdS}/\text{CFT}$ correspondence and large-$N$ limit in this procedure. 
\section{Gross-Neveu-Yukawa model in $\epsilon$ expansion and Gross-Neveu model in $1/N$ expansion\label{sect:gn}}
In this section we move from the pedagogical example of the $\varphi^4$ theory to the model with a more involved structure of the analytically continued spectrum. We consider two realizations of the same critical CFT. The first is the IR-stable critical point of the Gross-Neveu-Yukawa (GNY) model
\begin{equation}
\label{action-gny}
    S_{GNY} = \int d^d x\left(\bar{q}\slashed{\partial}q + \dfrac{1}{2}(\partial \sigma)^2 + g\bar{q}\sigma q + \dfrac{\lambda}{4!}\sigma^4\right),
\end{equation}
and the second is the Gross-Neveu (GN) model in the framework of $1/N$ expansion 
\begin{equation}
\label{action-gn}
    S_{GN} = \int d^dx \left(\bar{q}\slashed{\partial}q + \sigma\bar{q}q - \dfrac{N}{2g}\sigma^2\right).
\end{equation}
In both models $q_i(x)$ ($i = 1,\ldots, N$) is an $N$-component fermion field transforming under $SU(N)$ and $\sigma(x)$ is a scalar field.

The definition of singlet leading-twist operators in both models (see~\cite{Manashov:2025kgf} for the discussion on other types of operators) are the same
\begin{align}
    \mathcal{O}^{q}(s) =& -s\sum_{i = 1}^{N}\bar{q}_i(x)\gamma_{(\mu_1}\partial_{\mu_2}\ldots\partial_{\mu_s)}q_i(x) - \text{traces}, \nonumber\\
    \mathcal{O}^{\sigma}(s) =& \sigma(x)\partial_{\mu_1}\ldots\partial_{\mu_s}\sigma(x) - \text{traces}.
\label{GN-operators}
\end{align}
The difference arises in the scaling dimension of the operator $\mathcal{O}^{\sigma}(x)$ due to the different canonical scaling of $\sigma$ field ($\Delta^{GNY}_{\sigma, 0} = d/2 - 1$ in contrast with $\Delta^{GN}_{\sigma,0} = 1$). The operators~\eqref{GN-operators}, therefore, mix in the GNY model~\eqref{action-gny} and one has to diagonalize the anomalous-dimension matrix to access multiplicatively renormalizable operators. We denote the corresponding eigen-trajectories by $\Delta^{1,2}(s)$. In the GN model it is convenient to keep the notation $\Delta^q(s)$ and $\Delta^\sigma(s)$.

\begin{figure}[t]
\centering
\begin{tikzpicture}

\node (img1) at (0,0)
  {\includegraphics[width=0.45\textwidth]{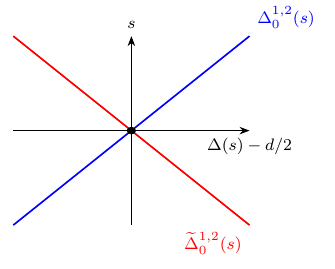}};
\node (img2) at (7,0)
  {\includegraphics[width=0.45\textwidth]{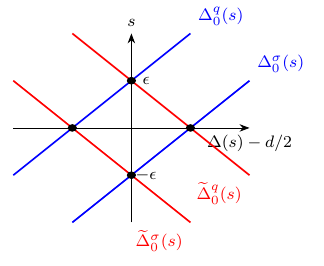}};

\node[anchor=north] at (img1.south) {(a)};
\node[anchor=north] at (img2.south) {(b)};

\end{tikzpicture}
\caption{Conformal Chew-Frautschi plot for the Gross-Neveu-Yukawa model in $\epsilon$ expansion (a) and Gross-Neveu model in $1/N$ expansion (b) around point $s = 0$. In Figure the scaling dimensions of the operators~\eqref{GN-operators} and their shadow scaling dimensions in free theory are depicted.}
\label{fig:regges}
\end{figure}

Let us focus on the point $s = 0$. In Fig.\ref{fig:regges} we plot the free conformal Regge trajectories, corresponding to the analytical continuation of~\eqref{GN-operators} and their shadows. We see that the situation in the GNY model is completely analogous to the example discussed earlier in section~\ref{sect:phi4}, and we will not repeat the analysis here (see~\cite{Manashov:2025kgf} for details). Switching to the GN model~\eqref{action-gn} reveals an interesting interplay between $\epsilon$ and $1/N$ expansions. Since $\epsilon$ is now an arbitrary parameter, the single point of intersection in Fig.~\ref{fig:regges} (a) splits into several crossings in Fig.~\ref{fig:regges} (b): at $s = \pm \epsilon$ and two at $s = 0$.       

Analyzing the result for the anomalous dimensions in $1/n$ order ($n = N\operatorname{tr}\mathds{1}, \mu = d/2$, $\eta_1$ -- critical exponent corresponding to the quark field)~\cite{Muta:1976js, Giombi:2017rhm}
\begin{align*}
\label{gn-res}
        \gamma_\sigma(s) &=-\dfrac{\eta_1}{n}\frac{2(2\mu-1)}{(\mu-1)}\left(1
        -\frac{\mu}{2\mu-1}\frac{\Gamma(\mu)}{\Gamma(s+\mu)}\frac{\Gamma(s+2-\mu)}{\Gamma(3-\mu)} \right) +O(1/n^2)\,,
        \\
        \gamma_q(s) &=\dfrac{\eta_1}{n}\left(1-
        \frac{\mu(\mu-1)}{(s+\mu-1)(s+\mu-2)}\left(1+\frac{\Gamma(2\mu-1) \Gamma(s+1)}{(\mu-1)\Gamma(2\mu-3+s)}\right)\right)+O(1/n^2)\,,   
\end{align*}
we see the consequences of these intersections, namely
\begin{align}
    \lim_{s \to -\epsilon}\gamma_{\sigma}(s) = \infty, && \lim_{s \to \epsilon}\gamma_{q}(s) = \infty. 
\end{align}
These singularities can be resummed using the same DLE-type formula~\eqref{resum-crit}. The point $s = 0$ is more interesting. Expressions~\eqref{gn-res} at $1/n$ order are finite. The only visible consequence is the mismatch between the limit $s \to 0$ and anomalous dimension of the operator $\sigma^2$
\begin{equation}
\label{gn-limit}
    \lim_{s \to 0}\gamma_{\sigma}(s) \neq \gamma_{\sigma^2}.
\end{equation}
This fact was noted in~\cite{Giombi:2017rhm}, but the reason behind such a discrepancy remained unclear.

Applying the logic described in section~\ref{sect:phi4} we can construct a defining equation for the exact trajectory near point $s =0$. Regularity of its coefficients leads to the definition of two regular functions
\begin{align}
    \omega_1(s) = \dfrac{1}{2}\big(\gamma_q(s) - \gamma_{\sigma}(s)\big), && \omega_2(s) = s\big(\gamma_{q}(s) + \gamma_{\sigma}(s)\big) + \gamma_{q}(s)\gamma_{\sigma}(s),
\end{align}
and the formula for resummation around $s = 0$ reads
\begin{equation}
\label{gn-resum}
    \gamma_{q,\sigma}(s) = \pm\omega_{1}(s) - s + \sqrt{s^2 + \omega_2(s) + \omega_1^2(s)}.
\end{equation}
Substituting~\eqref{gn-res} in~\eqref{gn-resum} and taking the limit $s \to 0$ one, indeed, removes the issue~\eqref{gn-limit}. 

Let us comment on why the usual perturbative expansion~\eqref{gn-res} produces the problem despite the absence of a singularity at $s = 0$ at order $1/N$. The first term in the function $\omega_2(s)$ is accompanied with $s$, so it vanishes in the limit $s \to 0$, unless there is a singularity in $\gamma_q(s) + \gamma_{\sigma}(s)$. Such a singularity is absent at leading order but does appear in the order $1/N^2$, which has to be taken into account to match the leading order of $\gamma_{q}(s)\gamma_{\sigma}(s)$. Indeed, using the result available for $\gamma_q(s)$ at the $1/N^2$ order~\cite{Manashov:2017xtt} we see the singularity in the result for the quark operator
\begin{equation}
    \gamma_q(s) = \left(-\dfrac{\eta_1^2}{n^2}\dfrac{2\mu(2\mu - 1)(2\mu - 3)}{(\mu - 2)^2}\times\dfrac{1}{s} + O\left(s^0\right)\right) + O\left(1/n^3\right).
\end{equation}
The $1/N^2$ calculation for $\gamma_{\sigma}(s)$ is not yet available. Luckily, the regularity of $\omega_1(s)$ forces the singularities of $\gamma_{\sigma}(s)$ and $\gamma_{q}(s)$ to coincide in all orders in the $1/N$ expansion. This fixes the needed behavior and makes the resummation consistent.
\section{Summary}

In these proceedings we have studied singularities of the anomalous dimensions of twist-2 operators as functions of spin. Our focus was on the singularities that share common features with those arising in the $s \to 0$ limit for flavor non-singlet quark operator in QCD.

Recent progress in formal conformal field theory~\cite{Caron-Huot:2022eqs} provides a systematic prescription for resumming such singular behavior. Although the functional form of the resummed result had been anticipated earlier from heuristic analyses based on the double-logarithmic equation~\cite{Velizhanin:2014dia}, the interpretation in terms of mixing of conformal Regge trajectories enriches the conceptual justification and provides better theoretical control. As a consequence, this framework can be extended to more involved situations (see Sec.~\ref{sect:gn} and~\cite{Manashov:2025kgf}), where straightforward generalizations of the original arguments are not obvious.

At the same time, extending this program to other types of singularities and to higher perturbative orders leads to new questions and requires further development. Already in~\cite{Caron-Huot:2022eqs} the authors show that methods based solely on the renormalization of local operators are not sufficient to reconstruct the full set of conformal Regge trajectories. Progress in this direction would be valuable, if only for a practical purpose, since it would provide additional constraints on the analytic structure of twist-2 anomalous dimensions. As an example we note the recent combination of BFKL-based predictions with the conformal approach, which unifies the consideration of the singularities at $s \to 1$ in gluon–gluon anomalous dimensions and $s \to 0$ in~\eqref{QCD-op-def} through the use of the so-called detector operators~\cite{Chang:2025zib}. The breakdown of resummation procedure based on DLE formula~\eqref{dle-qcd} at three loops suggests the need for further investigation of conformal Chew-Frautschi plot in QCD, if one wants to apply this logic at higher loop orders.  

\section*{Acknowledgments}

This work was supported by Deutsche Forschungsgemeinschaft (DFG) through the Research Unit FOR 2926, “Next Generation pQCD for Hadron Structure: Preparing for the EIC”, project number 40824754, and the ERC Advanced Grant 101095857 {\it Conformal-EIC}.
\newpage

\bibliography{refs.bib}
\bibliographystyle{JHEP}

\end{document}